\documentclass[prb,twocolumn,amsmath,amssymb,superscriptaddress,aps,10pt]{revtex4}

\usepackage{graphicx}
\usepackage{dcolumn}
\usepackage{bm}
\usepackage{gensymb}
\usepackage{amsmath}
\usepackage{color}
\usepackage{epsfig}
\usepackage{epstopdf}
\usepackage{soul}
\usepackage{booktabs} 
\usepackage{multirow}
\usepackage{microtype}
\usepackage{upgreek}
\usepackage{float}

\usepackage{lipsum}

\usepackage{hyperref}
\usepackage[utf8]{inputenc}
\usepackage[english]{babel}

\usepackage{xcolor}

\usepackage{tikz,xcolor,hyperref}

\definecolor{lime}{HTML}{A6CE39}
\DeclareRobustCommand{\orcidicon}{%
	\begin{tikzpicture}
	\draw[lime, fill=lime] (0,0) 
	circle [radius=0.16] 
	node[white] {{\fontfamily{qag}\selectfont \tiny ID}};
	\draw[white, fill=white] (-0.0625,0.095) 
	circle [radius=0.007];
	\end{tikzpicture}
	\hspace{-2mm}
}

\foreach \x in {A, ..., Z}{%
	\expandafter\xdef\csname orcid\x\endcsname{\noexpand\href{https://orcid.org/\csname orcidauthor\x\endcsname}{\noexpand\orcidicon}}
}

\begin{document}


\title{Wireless at the Nanoscale: Towards Magnetically Tunable Beam Steering}

\newcommand{\orcidauthorA}{0000-0001-6178-0061}
\newcommand{\orcidauthorB}{0000-0003-1742-9957}
\newcommand{\orcidauthorC}{0000-0002-2095-7008}

\author{William O. F. Carvalho\orcidA{}}
\email{williamofcarvalho@gmail.com}
\affiliation{National Institute of Telecommunications (Inatel), Santa Rita do Sapuca\'i, MG, 37540-000, Brazil}
\author{E. Moncada-Villa\orcidC{}}
\affiliation{Escuela de F\'isica, Universidad Pedag\'ogica y Tecnol\'ogica de Colombia, Av. Central del Norte 39-115, Tunja, Colombia}
\author{J. R. Mej\'ia-Salazar\orcidB{}}
\email{jrmejia3146@gmail.com}
\affiliation{National Institute of Telecommunications (Inatel), Santa Rita do Sapuca\'i, MG, 37540-000, Brazil}

\date{\today}

\begin{abstract}
Plasmonic wireless nanolinks hold great promise to overcome limitations from conventional metallic wires, namely narrow bandwidths, Ohmic losses, dispersion and cross-talking. However, current developments are limited to the wireless communication between two fixed points, i.e., a fixed transmitter with a fixed receiver, which in turn limits the energy efficiency and integration levels. In this work, we propose the use of magnetic fields for active tuning of the radiation beam steering of plasmonic nanoantennas. The physical principle behind our concept is shown with rigorous solution of the radiation pattern of a radiating dipolar source in the presence of magnetic fields. The results indicate that the proposed scheme is feasible with plasmonic nanoantennas made of ferromagnetic or noble metals in the presence of an applied magnetic field.
\end{abstract}

\maketitle

Nanoscale analogs of conventional radio frequency (RF) antennas, commonly called nanoantennas, have become a subject of considerable attention during the last decade.\cite{alu2007enhanced,Alu2008,alu2010wireless,Yang2016,Afridi2016,Cohen2017,Kullock2020} This idea was initially inspired by the possibility of using plasmonic dipoles as chip-scale optical/infrared wireless nanolinks,\cite{Alu2008,alu2010wireless} with the potential to overcome the major drawbacks of conventional nanoelectronic interconnects (e.g., copper wires), namely narrow bandwidths, Ohmic losses, dispersion and cross-talking.\cite{BrongersmaMark2010} Furthermore, owing to the unique confinement properties of plasmonic structures, these nanoantennas hold promise for improved levels of integration density and miniaturization of on-chip photonic devices.\cite{Wu2017,kim2020beam} However, there are two major limitations that need to be solved for this to become a reality. First, due to large optical wireless propagation losses, the transmitting ($T_{t}$) and receiving ($R_{r}$) nanoantennas are limited to be placed at distances in the order of the wavelength ($\lambda$). Second, in contrast to RF antennas, plasmonic wireless nanolinks are limited to the communication between two fixed $T_{t}$ and $R_{r}$, i.e., there are no mechanisms available to actively manipulate communication between a single $T_{t}$ with multiple $R_{r}$. Attempts to overcome the first limitation have mainly focused on geometric designs that optimize the values of directivity ($D_{r,t}$), which in turns optimize the ratio between received ($P_{r}$) and transmitted ($P_{t}$) power $\left(\frac{P_{r}}{P_{t}}\sim \frac{D_{r}D_{t}}{l^{2}}\textrm{, where $l$ represents the propagation length}\right)$.\cite{alu2010wireless} Indeed, plasmonic horn-like,\cite{Yang2016,Afridi2016} Yagi-Uda-like,\cite{Kullock2020} and plantenna\cite{Cohen2017} designs have demonstrated propagation distances of up to $3$ orders of magnitude larger than the plasmon wavelength. In contrast, finding mechanisms to dynamically manipulate the beam steering of plasmonic nanoantennas remains a challenge.

\begin{figure}[t]
    \centering
    \includegraphics[width=1\columnwidth]{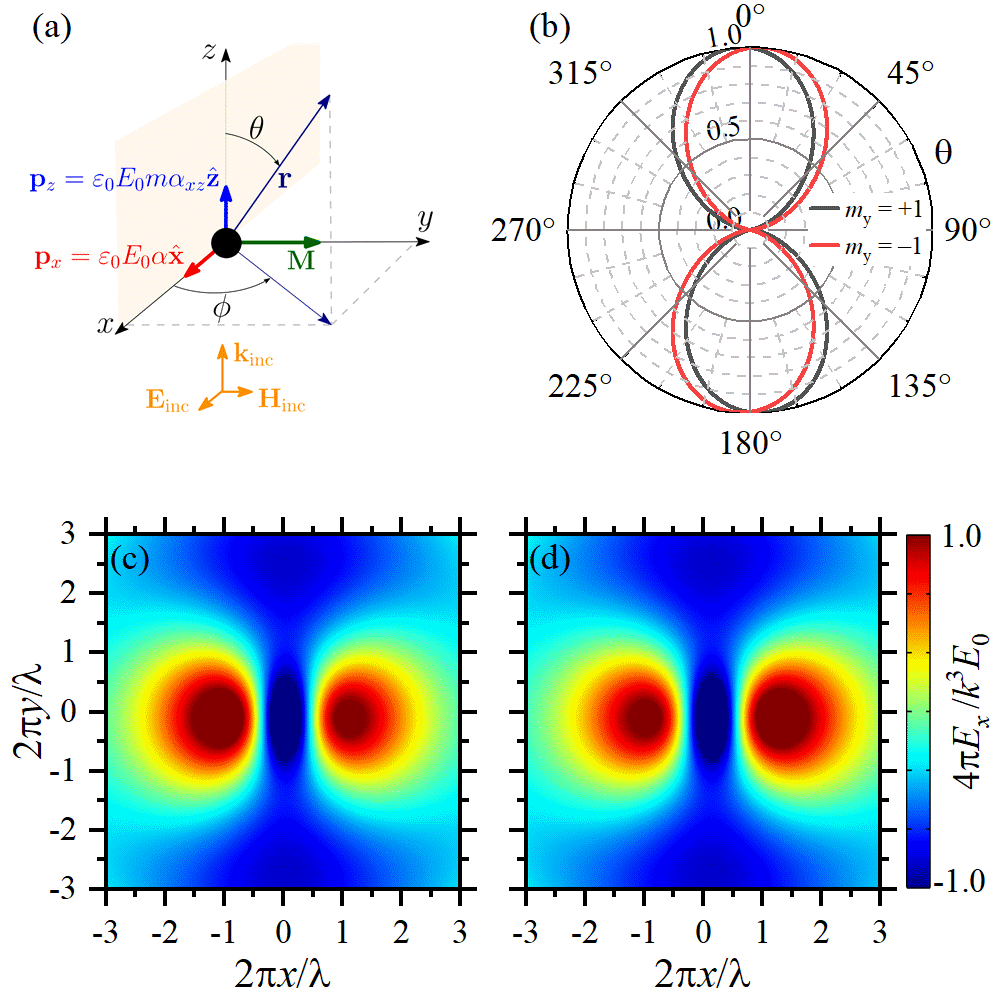}
    \caption{(a) Pictorial representation of a radiating electric dipole in the presence of an applied magnetic field (or having an intrinsic magnetization) $\textbf{M}$. (b) Radiation patterns for $\textbf{M}$ pointing along the positive ($+m$) and negative ($-m$) directions of the $y$-axis. Normalized $x$-component of the electric field (along the $xy$-plane) for (c) $+m$ and (d) $-m$ calculated at $z=\frac{\lambda}{2\pi}$. Calculations were made for a point dipole with radius $r_{d}=90$~nm, considering $\lambda=750$~nm, $\varepsilon=-14+i1.7$ and $\varepsilon_{xz}=-4.01-i2.27$.}
    \label{figs1}
\end{figure}

In this work, we examine the potential application of magnetic fields for active manipulation of the optical beam steering of plasmonic nanoantennas, i.e., the magnetically tunable beam steering. In contrast to one-dimensional and two-dimensional MO grating structures (with optical properties dictated by the corresponding diffraction orders), where the magnetic beam steering has been recently demonstrated,\cite{Tan2019,Faniayeu2020} we focused here on the active manipulation of the radiated field by a single magnetoplasmonic nanoantenna. Magnetic fields ($\mathbf{M}$) induce optical anisotropies, commonly called magneto-optical (MO) effects, that depend on the direction and sense of the applied field in relation to the direction of light propagation.\cite{visnovsky2018optics} Significantly, MO effects have been used to tune the localized surface plasmon resonance (LSPR) features of isolated nanoparticles,\cite{Sepulveda2010,Maccaferri2013,Maccaferri2016} as well as to control the light transmission, reflection, and absorption amplitudes by nanostructures made of noble metals,\cite{Sepulveda2010,kuttruff2021magneto} ferromagnetic metals,\cite{Maccaferri2013,Maccaferri2016} and hybrid systems.\cite{Yang2022}

Let us first discuss the radiative properties of an isolated point dipole ($\textbf{p}_{x}$) in the presence of a static magnetic field ($\textbf{M}$). The point dipole (of volume $V=\frac{4\pi}{3}r^{3}_{d}$) is excited by a monochromatic $x$-polarized incident plane wave, as illustrated in Fig.~\ref{figs1}(a). $\textbf{M}$ can be either an external magnetic field or an intrinsic magnetization, which causes an optical anisotropy described by the dielectric tensor
\begin{equation}
\label{perm-tensor}
\hat \varepsilon = 
\left( \begin{array}{ccc} 
\varepsilon  			  & 		0  				   & im\varepsilon_{xz}\\
0							   &	  \varepsilon    &    0   \\
-im\varepsilon_{xz}   &  0						&  \varepsilon
\end{array} \right),
\end{equation}
where $m=+1$ ($m=-1$) is used for $\textbf{M}$ pointing along the positive (negative) sense of the $y$-axis. Using the point dipole polarizability in the radiative limit, $\hat\alpha = \left(\hat{\mathbf 1} -i\frac{k^3}{6\pi}\hat{\alpha}_0\right)^{-1}\hat\alpha_0$, with $\hat\alpha_0 = 3V[\hat\varepsilon-\hat{\mathbf 1}][\hat\varepsilon+2\hat{\mathbf 1}]^{-1}$ for the static (non-radiative) term, we obtain a polarizability tensor with the following structure
\begin{equation}
\label{pol2}
\hat \alpha = 
\left( \begin{array}{ccc} 
\alpha  			  & 		0  				   & m\alpha_{xz}\\
0							   &	  \alpha    &    0   \\
-m\alpha_{xz}   &  0						&  \alpha
\end{array} \right).
\end{equation}

\begin{figure}[t]
    \centering
    \includegraphics[width=1\columnwidth]{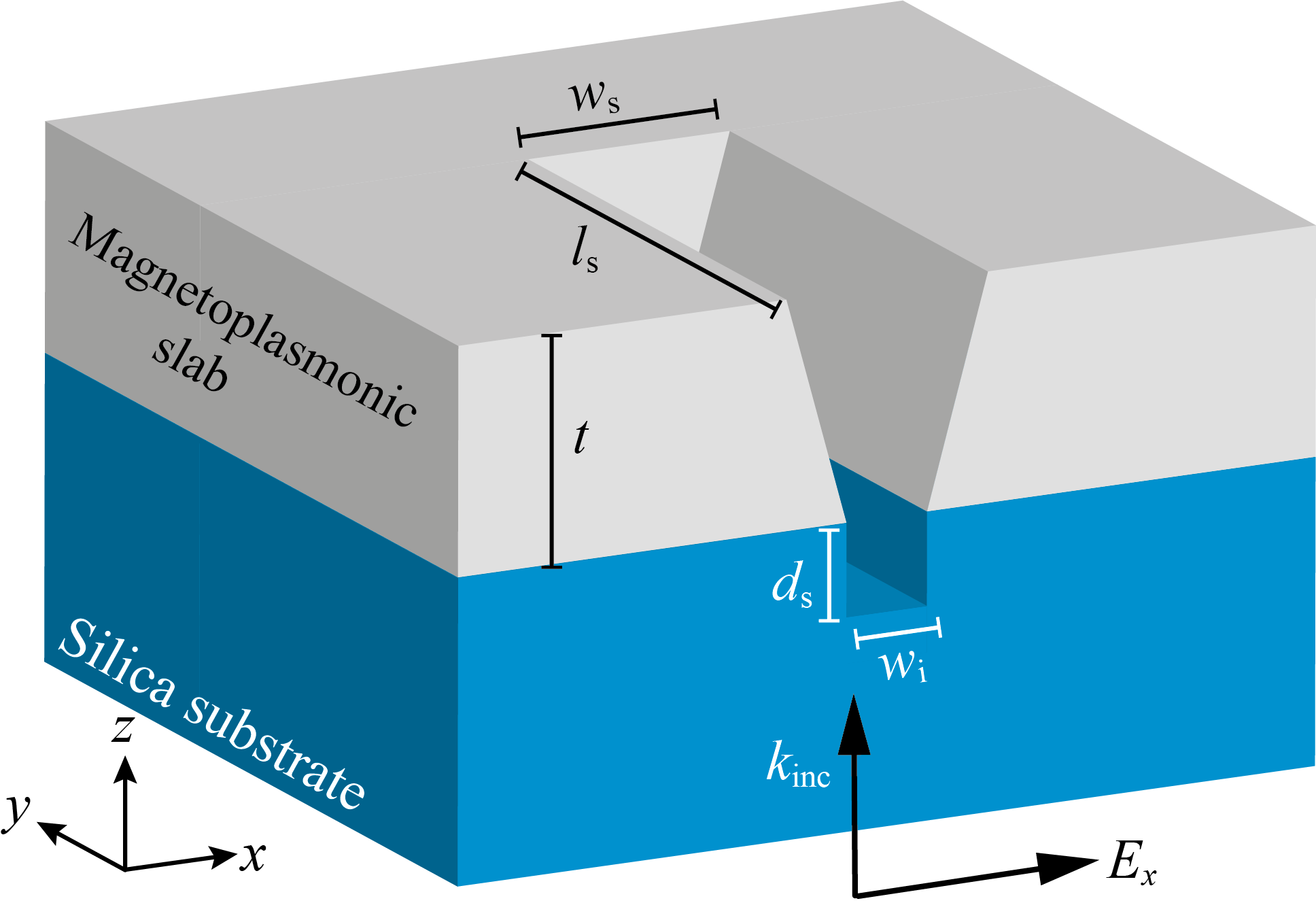}
    \caption{Schematic representation of the magnetoplasmonic nanoantenna under study. The incident electric field points along the $x$-direction, in agreement with analytical results in Fig.~\ref{figs1}.}
    \label{figs2}
\end{figure}

Calculating the induced dipolar moment $\mathbf{p} =\varepsilon_0\hat\alpha\cdot\mathbf{E}_{\rm inc}=\varepsilon~E_0\alpha(\hat{\mathbf x}-m\eta\hat{\mathbf{z}})$, with $\varepsilon_0$ for the vacuum dielectric permittivity and $\mathbf E_{\rm inc}=E_0\hat{\mathbf x}$ and $\eta=\alpha_{xz}/\alpha$, we note that the interaction with $\textbf{M}$ induces a dipolar moment along the $z$-axis ($\mathbf p_z$),\cite{Maccaferri2013} as depicted in Fig.~\ref{figs1}(a). The radiated electric and magnetic fields, at a generic position $\mathbf r$, are given by \cite{Novotny2012,Ott2018}
\begin{eqnarray}\label{Efield}
	\mathbf{E}_{i}^{\rm sca}(\mathbf r) &=& \frac{k^2}{\varepsilon_0} \hat{\mathbf G}^{\rm E} (\mathbf r,\mathbf r_i) \cdot\mathbf p_i,\\  \label{Hfield}
	\mathbf{H}_{i}^{\rm sca}(\mathbf r) &=& \frac{k^2}{\varepsilon_0} \hat{\mathbf G}^{\rm H}(\mathbf r,\mathbf r_i) \cdot \mathbf p_i,
\end{eqnarray}
where $k=\omega/c$ is the wavevector magnitude of the radiated fields, $\omega$ is the angular frequency, $c$ is the speed of light in vacuum, and $\mathbf r_i$ is the position of the particle. $\hat{\mathbf G}^{\rm E} (\mathbf r,\mathbf r_i)$ and $\hat{\mathbf G}^{\rm H} (\mathbf r,\mathbf r_i)$ represent the dyadic Green's functions for the radiated electric and magnetic fields, respectively, written as
\begin{figure}[t]
    \centering
    \includegraphics[width=1\columnwidth]{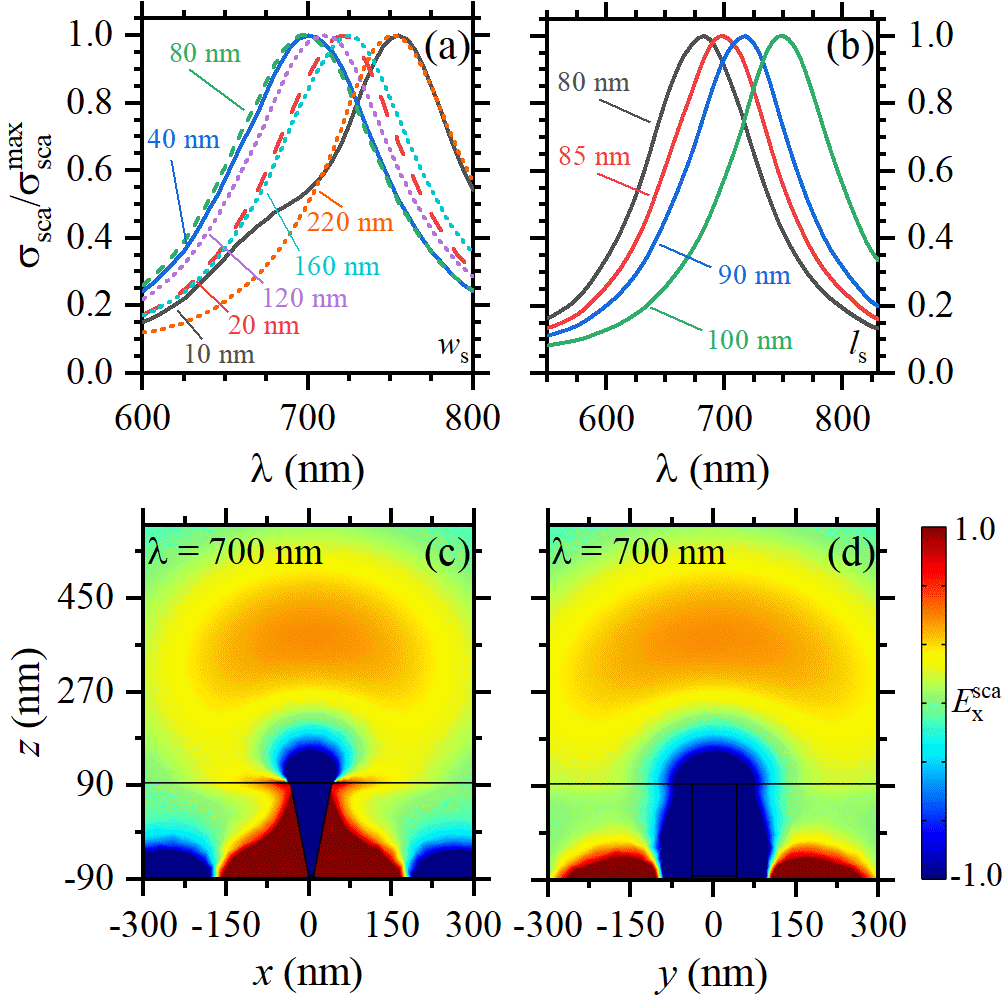}
    \caption{Normalized scattering cross-section spectrum, as function of $\lambda$, for different (a) $w_{\rm s}$ (with $l_{\rm s}=85$~nm) and (b) $l_{\rm s}$ (with $w_{\rm s}=80$~nm). Normalized $E^{\rm sca}_{x}$ near-field along the (c) $xz$-plane and (d) $yz$-plane for the system with $w_{\rm s}=80$~nm and $l_{\rm s}=85$~nm, for $\lambda=700$~nm.}
    \label{figs3}
\end{figure}
\begin{eqnarray}\label{green}
 \hat{\mathbf G}^{\rm E}(\mathbf r,\mathbf r_i) & = & \frac{e^{ikR_i}}{4\pi R_i } \left[ \left(1+\frac{ikR_i-1}{(kR_i)^2}\right)\hat {\mathbf 1} \right. 
 \\\nonumber && + \left. \left( \frac{3-3ikR_i-(kR_i)^2}{(kR_i)^2}\right) \hat{\mathbf R}_i\otimes \hat{\mathbf R}_i\right] \\
 \hat{\mathbf G}^{\rm H}(\mathbf r,\mathbf r_i) & = & \frac{e^{ikR_i}}{4\pi Z_0R_i} \left(1+\frac{i}{kR_i}\right) \hat{\mathbf R}_i\times\hat{\mathbf 1},
 \end{eqnarray}
where $\mathbf R_i=\mathbf r - \mathbf r_i$, $R_i=|\mathbf{R}_i|$, $\hat {\mathbf{R}}_i=\mathbf{R}_i/R_i$, $\otimes$ denotes a dyadic product, and $\hat{\mathbf R}_i\times\hat{\mathbf 1}$ represents the matrix generated by the cross-product of $\hat{\mathbf R}_i$ with each column vector of $\hat{\mathbf 1}$. $Z_0=\sqrt{\mu_0/\varepsilon_0}$ is used for the vacuum impedance. Without lack of generality, we can assume the dipole placed at the origin of the coordinate system and, therefore, calculate the Poynting vector $\mathbf S _k(\mathbf r)= 2{\rm Re}\{  \mathbf{E}(\mathbf r) \times \mathbf{H}^*(\mathbf r)  \}$. The radially radiated power $S_{k,{\rm rad}}(\mathbf r)=\mathbf S_k(\mathbf r)\cdot \hat r $ is given by
\begin{eqnarray}\nonumber
S_{k,{\rm rad}}(\mathbf r)&=& \frac{1}{2Z_0}\left( \frac{k^3|E_0||\alpha|}{4\pi}  \right)^2 \left[\frac{ f(\theta,\phi)}{(kr)^2}-\frac{g(\theta,\phi)}{(kr)^5} \right] ,\\ \label{poynting}
\end{eqnarray}
where
\begin{eqnarray}\nonumber
f(\theta,\phi)&=&3+2|\eta|^2+(1-2|\eta|^2)\cos 2\theta +2m\sin 2\theta\\ \nonumber
&& \times(2{\rm Re}\{\eta\}\cos \phi -m|\eta|^2  \sin\phi)  \\ \label{poyn_ang} 
&& - 2\sin^2\theta (\cos 2\phi - m {\rm Re}\{	\eta\} \sin 2\phi),\\
g(\theta,\phi)&=&2m{\rm Im}\{\eta\}\sin^2\theta\sin 2\phi.
\end{eqnarray}
From these analytical results, we conclude that the direction of the radiated beam (from a dipolar source) can be actively tuned by the sense (along the $y$-axis) and magnitude of $\textbf{M}$. Indeed, we show this last result numerically in Fig.~\ref{figs1}(b), where the radiation pattern (at the $xz$-plane, i.e., $\phi=0^{\circ}$) presented for $m=\pm 1$. These numerical results are for a point dipole with radius $r_{d}=90$~nm, considering $\lambda=750$~nm, $\varepsilon=-14+i1.7$ and $\varepsilon_{xz}=-4.01-i2.27$. The magnetically tunable beam steering in Fig.~\ref{figs1}(b) can be understood from the spatial profile of the $x$-component of the radiated electric field ($E_{x}^{\rm sca}$), which can be written as
\begin{eqnarray}\nonumber
E_{x}^{\rm sca}(r,\theta,\phi)&=&\frac{k^3E_0}{4\pi} \frac{\alpha e^{ikr}}{(kr)^3}\{-1+kr(i+kr) \\ \nonumber 
&&  +[-3+ kr(3i+kr)]\sin\theta \cos\phi\times \\ \label{Exfield}
&&(m\eta \cos\theta  - \sin\theta\cos\phi) \} .
\end{eqnarray}
\begin{figure}[t]
    \centering
    \includegraphics[width=1\columnwidth]{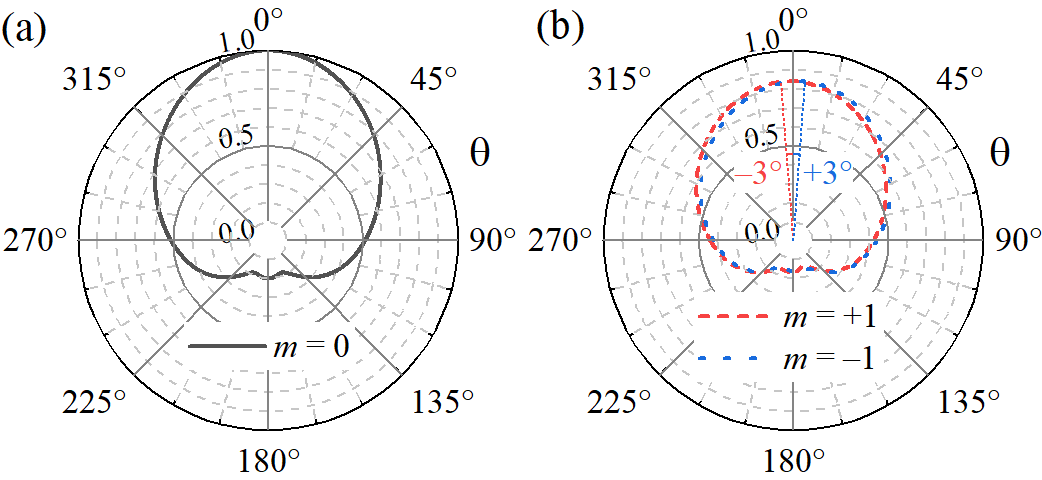}
    \caption{Radiation pattern of the system in Fig.~\ref{figs3} for (a) $m=0$ and (b) $m=\pm1$.}
    \label{figs4}
\end{figure}

\begin{figure}[t]
    \centering
    \includegraphics[width=1\columnwidth]{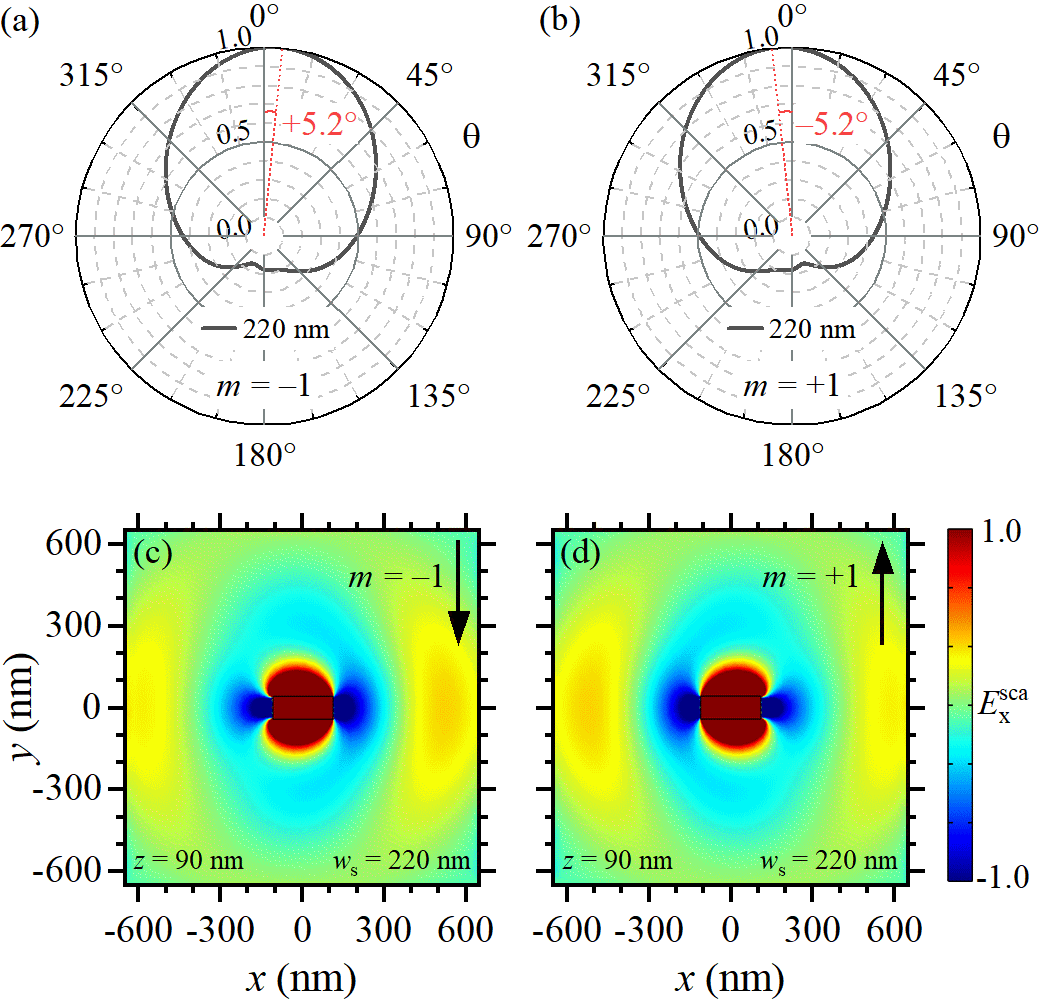}
    \caption{Normalized radiation patterns for the system with $w_{\rm s} = 220$~nm and $l_{\rm s} = 85$~nm for (a) $m=-1$ and (b) $m=+1$. Red dotted lines are used to indicate the radiated beam direction for each $m$.}
    \label{figs5}
\end{figure}

Numerical results of Eq.~(\ref{Exfield}) for $m=\pm 1$ are shown in Figs.~\ref{figs1}(c)-(d) along the $xy$-plane, calculated at $z=\frac{\lambda}{2\pi}$. The asymmetry of $E_{x}^{\rm sca}$ for $m=1$ and $m=-1$, around the axis $x=0$, explains the change in the tilt direction of the radiation beam when the magnetic field is reversed. Since radiative contributions of plasmonic nanoantennas are dominated by dipole resonances,\cite{giannini2011plasmonic} the physical principle demonstrated above can be exploited for magnetically tunable beam steering of plasmonic nanoantennas, as it will be shown below. To develop our idea, we use a well-known nanostructure, namely horn-like aperture plasmonic nanoantenna, which has been used for optical wireless communications\cite{kim2014babinet} and plasmonic biosensing applications.\cite{Barulin2022} The system can be considered built by a noble metal (or ferromagnetic metal) film of thickness $t$, with a horn-like nanoaperture, placed on a conventional SiO$_{2}$ glass substrate, as illustrated in Fig.~\ref{figs2}. The nanoaperture has a rectangular cross-section with a fixed length $l_{s}$, whereas the width changes linearly from a length $w_{s}$ at the top to a length $w_{i}$ at the bottom of the structure. A rectangular cavity, of width $w_{i}$ and depth $d_{s}$, is used in the substrate (at the bottom of the nanoantenna) to optimize the resonance features of the system.\cite{kim2014babinet} Following the analytical model described in Eqs. (\ref{perm-tensor})-(\ref{Exfield}), we consider the nanoantenna fed by an $x$-polarized incident plane wave, as illustrated in Fig.~\ref{figs2}. Numerical results are obtained with the finite element method (FEM) using the commercial software COMSOL Multiphysics\textsuperscript{\textregistered}. Refined irregular-mesh-sizes were used for improved precision, with a mesh-size of $\lambda/20$ near the aperture and $\lambda/10$ for the rest of the structure. Perfectly matched layers (PMLs) are used to avoid undesired numerical reflections. The wavelength ($\lambda$) is swept in the range from $550$~nm to $830$~nm, with steps of $1$~nm. To discuss our results in the context of realistic material parameters, we start considering the magnetoplasmonic slab made of a metallic ferromagnetic cobalt-silver alloy (Co$_6$Ag$_{94}$).\cite{wang1999study,Chai2021,Shpetnyi2021} The resonant wavelength of the structure (maximum $\sigma_{\rm sca}$) is studied as a function of $w_{s}$ (with $l_{s}=85$~nm) and $l_{s}$ (with $w_{s}=80$~nm) in Figs.~\ref{figs3}(a)-(b), respectively. Since focused ion beam (FIB) milling is the most widely used technique to fabricate this type of structure, we used $w_{i}=10$~nm in agreement with the current minimum size limitation of the technique.\cite{kim2014babinet} 

Focusing our attention on the structure with $w_{s}=80$~nm and $l_{s}=85$~nm, with a resonant wavelength $\lambda=700$~nm, we plot the normalized $x$-component of the scattered field ($E_{x}^{\rm sca}$) along the $xz$- and $yz$-planes in Figs.~\ref{figs3}(c)~and~\ref{figs3}(d), respectively. These calculations were performed for a demagnetized Co$_6$Ag$_{94}$ film ($\textbf{M}=\textbf{0}$), i.e., for $m=0$ in Eq. (\ref{perm-tensor}), where we used $\varepsilon$ and $\varepsilon_{xz}$ from the available experimental data.\cite{wang1999study} In particular, we used $\varepsilon=-12.45+i1.766$ and $\varepsilon_{xz}=-3.34721-i2.75$ for $\lambda=700$~nm. The corresponding radiation patterns for $m=0$ and $m=\pm1$ are shown in Figs.~\ref{figs4}(a)-(b). The radiation pattern is slightly tilted from $0^{\circ}$ to $-3^{\circ}$ ($+3^{\circ}$) when $m$ changes from $0$ to $+1$ ($-1$). In contrast to Fig.~\ref{figs4}(a) (for $m=0$), where $D=1.0$, we noticed lower values of directivity ($D=0.84$) for $m=\pm1$ in Fig.~\ref{figs4}(b), which are explained by small magnetically tuned resonance shifts. Indeed, magnetic tuning of plasmonic resonances is the most widely used principle for magnetoplasmonic biosensing applications.\cite{mejia2018plasmonic,Rizal2021} Importantly, as shown in the Supplemental Material (SM) file, the magnetically tunable beam steering demonstrated in Fig.~\ref{figs4} can only be obtained for $\textbf{M}$ placed perpendicular (transverse) to the plane formed by $\textbf{p}_{x}$ and $\textbf{k}_{\rm inc}$ (illustrated in Fig.~\ref{figs1}).

\begin{figure}[t]
    \centering
    \includegraphics[width=1\columnwidth]{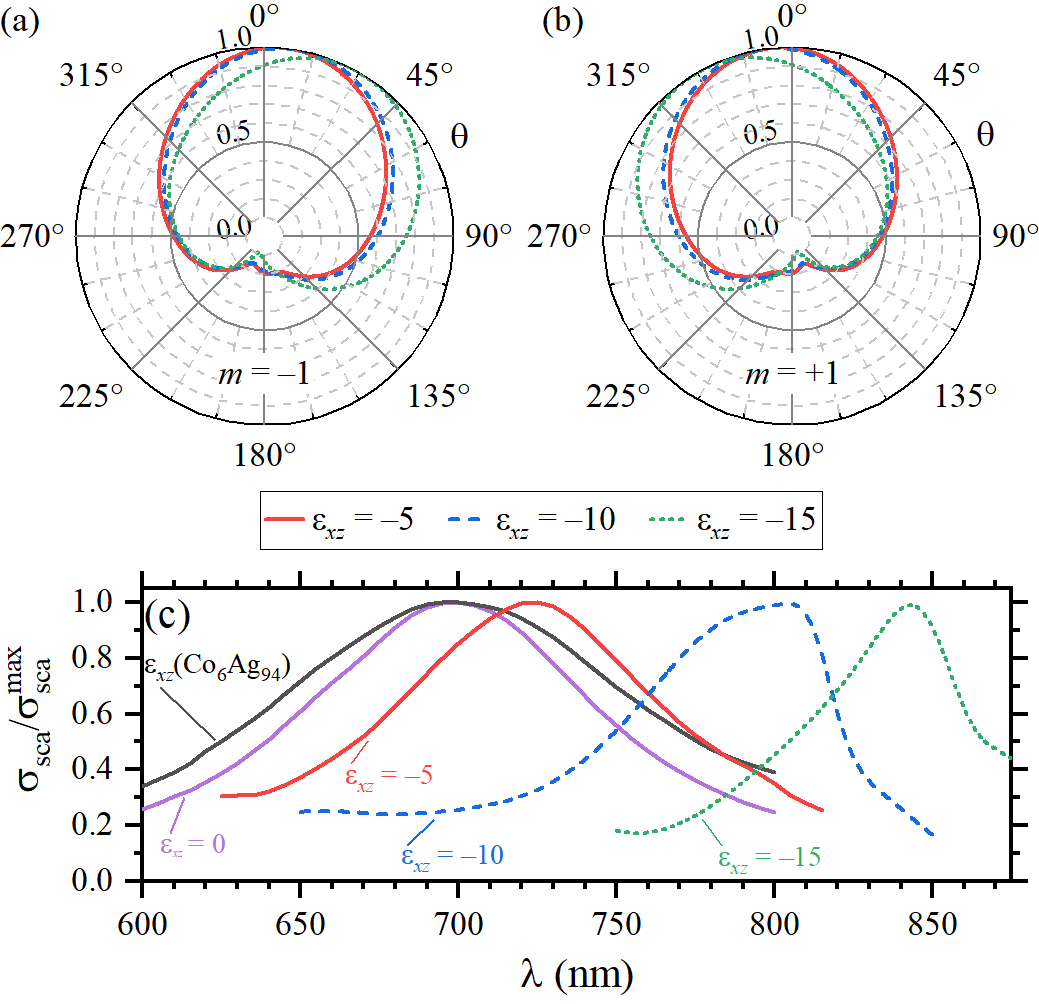}
    \caption{Normalized radiation patterns for (a) $m=-1$ and (b) $m=1$ associated to different values of $\varepsilon_{xz}$. Solid, dashed, and dotted lines are for $\varepsilon_{xz}=-5$, $\varepsilon_{xz}=-10$, and $\varepsilon_{xz}=-15$, respectively. (c) Normalized scattering cross-sections for $\varepsilon_{xz}=0$, $\varepsilon_{xz}=-5$, $\varepsilon_{xz}=-10$, and $\varepsilon_{xz}=-15$ are comparatively shown with the values for the Co$_6$Ag$_{94}$ material.}
    \label{figs6}
\end{figure}

Using the geometric tuning of the resonant wavelength, we also considered regions where the Co$_6$Ag$_{94}$ film exhibits higher MO activity. In particular, we studied the system at $\lambda=725$~nm ($\varepsilon_{xz}=-3.681-i2.174$) and $\lambda=750$~nm ($\varepsilon_{xz}=-4.016-i2.275$), corresponding to $w_{\rm s} = 160$~nm and $w_{\rm s} = 220$~nm (shown in Fig.~\ref{figs3}(a)), where the maximum changes in the beam steering were $\theta_{\pm} = \mp 4.6^{\circ}$ and $\theta_{\pm} = \mp 5.2^{\circ}$ (in relation to $\theta =0^{\circ}$), respectively. The subindex $\pm$ is used to indicate the sign of $m$ in the calculations. The normalized radiation patterns for $w_{\rm s} = 220$~nm are shown in Figs.~\ref{figs5}(a)-(b) for $m=-1$ and $m=+1$, respectively. Furthermore, the normalized $E_x^{\rm sca}$ near-fields associated to $m=\mp1$ are shown in Figs.~\ref{figs5}(c)-(d). We should remark that numerical results for the magnetoplasmonic nanoantenna in Fig.~\ref{figs5} are in excellent qualitative agreement with analytical results in Fig.~\ref{figs1} (for an isolated dipole). Significantly, the increase in beam steering with $\varepsilon_{xz}$ can be explained directly through Eqs.~(\ref{poynting})-(\ref{Exfield}).

Since the formalism developed in this work is not limited to the use of ferromagnetic materials, but to radiating electric dipoles in the presence of applied magnetic fields, we will consider hypothetical lossless values for $\varepsilon_{xz}$. It is worth mentioning that the value of $\varepsilon_{xz}$ increases with $\left|\mathbf{M}\right|$.\cite{MoncadaVilla2015} Therefore, these hypothetical scenarios can represent the case of a radiating plasmonic nanoantenna in the presence of high $\mathbf{M}$ values. Although results can be calculated for magnetizable or non-magnetizable plasmonic materials (like, for example, Au or Ag),\cite{Sepulveda2010} we continue using the diagonal permittivity values ($\varepsilon$) of Co$_6$Ag$_{94}$ for comparative purposes. Normalized numerical results for the radiation patterns, associated to $m=-1$ and $m=+1$, are shown in Figs.~\ref{figs6}(a)-(b), respectively. As expected from Eqs.~(\ref{poynting})-(\ref{Exfield}), the beam steering angle ($\theta$) increases with $\varepsilon_{xz}$. In particular, we obtained $\theta_{\pm}=\mp 7.2^{\circ}$, $\theta_{\pm}=\mp 10.8^{\circ}$ and $\theta_{\pm}=\mp 30.6^{\circ}$ for $\varepsilon_{xz}=-5$, $\varepsilon_{xz}=-10$ and $\varepsilon_{xz}=-15$, respectively. Radiation patterns were calculated at the corresponding plasmon resonances for each $\varepsilon_{xz}$. Fig.~\ref{figs6}(c) shows that, in contrast to ferromagnetic materials (inducing negligibly small resonance shifts), the use of high magnetic fields (large $\varepsilon_{xz}$) will strongly alter the resonant wavelength. These last results indicate a way to develop a new generation of magnetically tunable optical wireless nanolinks. In fact, our idea allows the possibility of using a single transmitting plasmonic nanoantenna to communicate with multiple receiving plasmonic nanoantennas (with different resonant wavelengths), located in front, which in turn enables further integration by reducing the number of circuit-elements.

In conclusion, we have theoretically shown that magnetic fields can be used to dynamically tune the direction of the beam radiated by plasmonic nanoantennas (which we called, in general, magnetoplasmonic nanoantennas). It was analytically and numerically demonstrated (see the SM) that the applied magnetic field should be placed perpendicular to the plane formed by the dipole ($\textbf{p}_{x}$) and the incident wavevector ($\textbf{k}_{i}$). Our concept provides a new paradigm for active beam steering of wireless signals at nano- and micro-scale levels. From the applications point of view, the communication of a single $T_{t}$ with multiple $R_{r}$ will reduce the number of on-chip/inter-chip photonic elements, allowing for improved energy efficiency and higher integration levels. Besides, the physical principle in this work lies on the optical radiation properties of electric dipolar sources in the presence of a magnetic field. Hence, the proposed scheme is not limited to a specific plasmonic nanoantenna design, i.e., it can be applied to any plasmonic nanoantenna.

\acknowledgments
Partial financial support was received from RNP, with resources from MCTIC, Grant No. 01245.010604/2020-14, under the Brazil 6G project of the Radiocommunication Reference Center (Centro de Refer\^{e}ncia em Radiocomunica\c{c}\~oes - CRR) of the National Institute of Telecommunications (Instituto Nacional de Telecomunica\c{c}\~oes - Inatel), Brazil. We also acknowledge financial support from the Brazilian agencies Coordena\c{c}\~ao de Aperfei\c{c}oamento de Pessoal de N\'ivel Superior - Brasil (CAPES) - Finance Code 001, the National Council for Scientific and Technological Development-CNPq (314671/2021-8,403827/2021-3) and FAPESP (2021/06946-0).

\end{document}